%% file: main.tex
\documentclass[sigconf,nonacm]{acmart}
\acmSubmissionID{1644}
\renewcommand\footnotetextcopyrightpermission[1]{}
\usepackage{tikz}
\usepackage{amsmath}
\usepackage{graphicx}
\usepackage{booktabs}
\usepackage{stfloats}
\usepackage{multirow}
\usepackage{todonotes}
\usepackage{subcaption}
\usepackage{xspace}
\usepackage[normalem]{ulem}
\graphicspath{{figures/}}
\newlength{\panelH}
\newcommand{\projectname}{Hazel}
\newcommand{\projectnameSpace}{Hazel\xspace}
\newcommand{\ourTree}{HMT\xspace}

\definecolor{blueish}{RGB}{0,113,197}
\newcommand*\negcircnum[1]{\tikz[baseline=(char.base)]{%
    \node[white,shape=circle,fill=blueish,draw,inner sep=1pt] (char) {\color{white}\normalfont\sffamily #1};}}

\definecolor{contribution_blue}{HTML}{026EC0}
\newcommand{\blueblock}{%
  \textcolor{contribution_blue}{\rule{2ex}{2ex}}%
}
\newcommand{\intext}[1]{%
  \raisebox{-0.2em}{\includegraphics[height=1em]{#1}}%
}

\usepackage{enumitem}
\setlist[enumerate]{%
  noitemsep,   
  topsep=5pt,  
  leftmargin=22pt, 
}

\begin{document}


\title{Hazel: Secure and Efficient Disaggregated Storage} 

\author{Marcin Chrapek}
\email{marcin.chrapek@inf.ethz.ch}
\affiliation{%
  \institution{ETH Zurich}
  \city{Zurich}
  \country{Switzerland}
}

\author{Meni Orenbach}
\affiliation{%
  \institution{NVIDIA}
  \city{Santa Clara}
  \country{USA}}

\author{Ahmad Atamli}
\affiliation{%
  \institution{NVIDIA}
  \city{Santa Clara}
  \country{USA}}
\affiliation{
  \institution{University of Southampton}
  \city{Southampton}
  \country{UK}}

\author{Marcin Copik}
\affiliation{%
  \institution{ETH Zurich}
  \city{Zurich}
  \country{Switzerland}
}

\author{Mikhail Khalilov}
\affiliation{%
  \institution{ETH Zurich}
  \city{Zurich}
  \country{Switzerland}
}

\author{Fritz Alder}
\affiliation{%
  \institution{NVIDIA}
  \city{Santa Clara}
  \country{USA}}

\author{Torsten Hoefler}
\affiliation{%
  \institution{ETH Zurich}
  \city{Zurich}
  \country{Switzerland}
}

\input{0_abstract}

\maketitle

\input{1_intro}

\input{3_threat_model}
\input{3_motivation}
\input{4_design}
\input{5_implementation}
\input{6_evaluation}
\input{7_discussion}

\section*{Acknowledgments: } We thank NVIDIA for providing computational resources and hosting MC's internship, where \projectnameSpace was developed. This research also obtained partial funding from the ``UrbanTwin: An urban digital twin for climate action: Assessing policies and solutions for energy, water and infrastructure'' project, funded by the ETH-Domain Joint Initiative program in the Strategic Area Energy, Climate and Sustainable Environment, a donation from Intel Corporation, a grant (agreement PSAP, No. 101002047) from the European Research Council (ERC) under the European Union’s Horizon 2020 program, and a grant (grant agreement No. 101167904) from the European Research Council (ERC) under the European Union's Horizon 2023 research and innovation program.

\bibliographystyle{ACM-Reference-Format}
\bibliography{references}

\end{document}

%% file: 0_abstract.tex
\begin{abstract}


Disaggregated storage with NVMe-over-Fabrics (NVMe-oF) has emerged as the standard solution in modern supercomputers and data center clusters, achieving superior performance, resource utilization, and power efficiency. Simultaneously, confidential computing (CC) is becoming the de facto security paradigm, enforcing stronger isolation and protection for sensitive workloads. However, securing state-of-the-art storage with traditional CC methods struggles to scale and compromises performance or security.
To address these issues, we introduce \projectname, a storage management system that extends the NVMe-oF protocol capabilities and adheres to the CC threat model, providing confidentiality, integrity, and freshness guarantees. \projectnameSpace offers an appropriate control path with novel concepts such as counter-leasing. 
\projectnameSpace also optimizes data path performance by leveraging NVMe metadata and introducing a new disaggregated Hazel Merkle Tree (\ourTree), all while remaining compatible with NVMe-oF. For additional efficiency, \projectnameSpace also supports offloading to CC-capable smart NIC accelerators. We prototype \projectnameSpace on an NVIDIA BlueField-3 and demonstrate that it can achieve as little as 1-2\% performance degradation for synthetic patterns, AI training, IO500, and YCSB.

\end{abstract}

%% file: 1_intro.tex
\section{Introduction}

Storage is the backbone of modern supercomputers and data center clusters, affecting the performance of applications~\cite{luttgauSurveyStorageSystems2018, zhao2022understanding}, microservices~\cite{maoPerformanceStudyVM2012}, serverless platforms~\cite{mcgrathServerlessComputingDesign2017a}, and the X-as-a-Service (XaaS) ecosystem~\cite{duanEverythingServiceXaaS2015}. 
The cost-effectiveness of storage has been improved by disaggregation~\cite{klimovicFlashStorageDisaggregation2016}, which decouples it from compute, enabling independent scaling and more efficient resource utilization. 
Disaggregated storage enabled by NVMe over Fabrics (NVMe-oF)~\cite{NVMeFabricsSpecification2021} achieves microsecond-level latencies and millions of IOPS by leveraging remote direct memory access (RDMA)~\cite{kaliaDesignGuidelinesHigh2016}, smart NIC (sNIC) offloading~\cite{minGimbalEnablingMultitenant2021, kimLineFSEfficientSmartNIC2021}, and 200-400 Gbit networks.

We also observe the convergence of HPC and the cloud~\cite{hoefler2022convergence}, as domains such as artificial intelligence, finance, and healthcare increasingly leverage the cost and scalability benefits of cloud service providers (CSPs). 
With these data-sensitive industries requiring strict workload protection, security became one of the most critical business determinants for CSPs~\cite{vossEuropeanUnionData2017, goldfarbShiftsPrivacyConcerns2012, petrescuAnalyzingAnalyticsData2018}. 
Combined requirements for security and performance led to the emergence of confidential HPC (cHPC)~\cite{chrapekHEARHomomorphicallyEncrypted2023}.
cHPC relies on Confidential Computing (CC)~\cite{russinovichConfidentialComputingElevating2023, mulliganConfidentialComputingBrave2021}, which emerged as a de facto security standard for the future of data centers.
In CC, users (tenants) can verifiably attest that their programs run in secure, isolated runtimes called trusted execution environments (TEEs)~\cite{costanIntelSGXExplained2016b,pintoDemystifyingArmTrustZone2019a,chengIntelTDXDemystified2024a,sev2020strengthening,liDesignVerificationArm2022a}.
TEEs reduce the risk of data leakage by limiting access, even from privileged entities such as cluster administrators. 

\begin{figure}[t] 
    \centering
\includegraphics[width=\columnwidth]{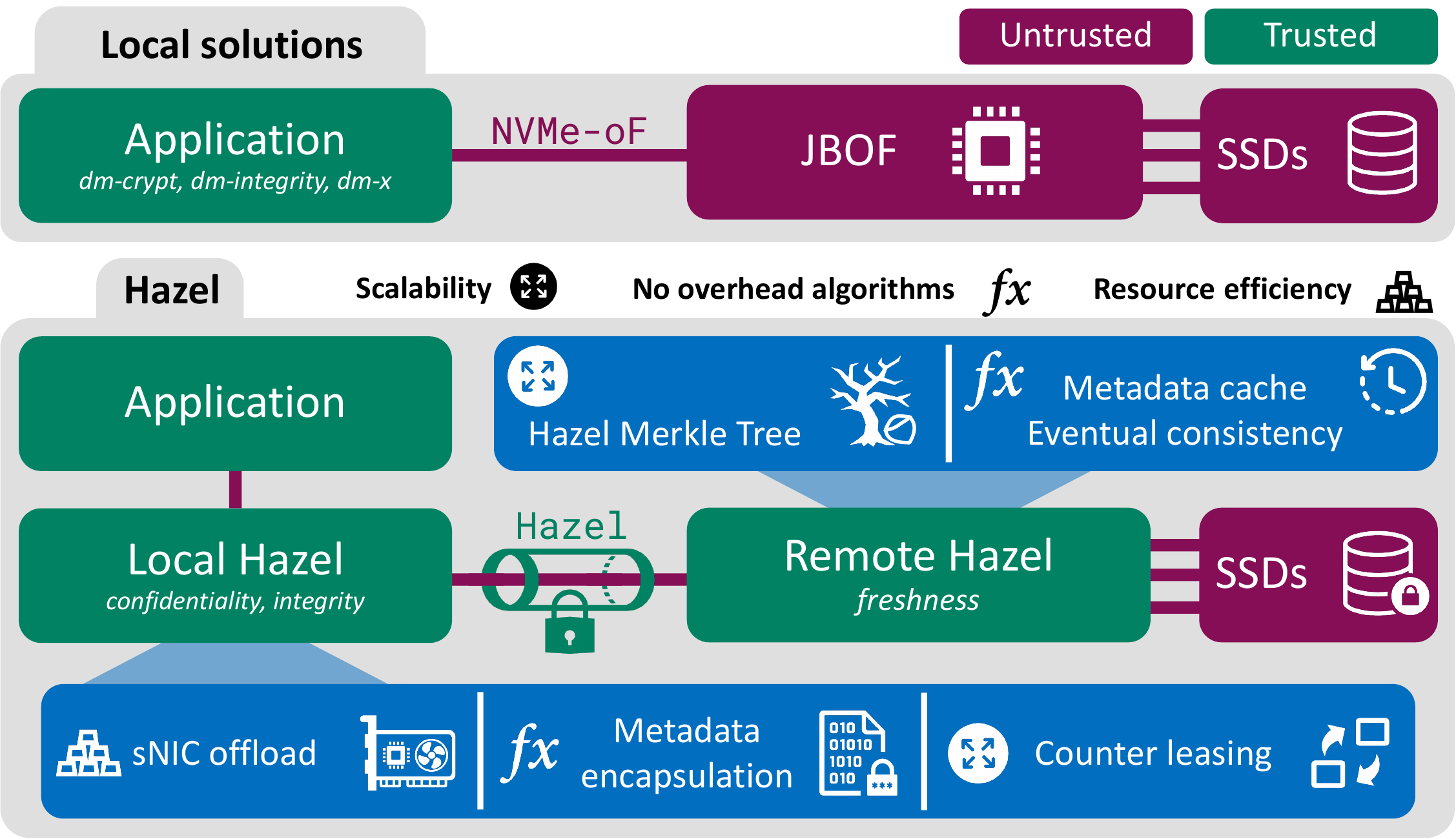} 
    \caption{An overview of the locally deployed methods and \projectnameSpace with its novel contributions~\blueblock\ in scalability~\intext{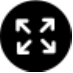}, algorithms~\intext{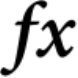}, and resource efficiency~\intext{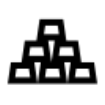}.}
    \vspace{-1em}
    \label{fig:child_poster}
\end{figure}

Secure storage is essential for CC systems~\cite{IntelTrustDomain}.
However, CC changes the typical threat model for storage.
Existing cloud storage solutions offered by CSPs provide confidentiality only through basic disk encryption with constant initialization vectors (IVs). 
Such protection is insufficient for CC, which requires unique IVs and two additional properties~\cite{badramsp25}: integrity (data is unmodified), and freshness (data is not outdated). 
%
%
User-based solutions designed for local deployments, such as \textit{dm-x}~\cite{chakrabortiDmxProtectingVolumelevel2017}, \textit{dm-crypt}, and \textit{dm-integrity}, offer some of these guarantees.
However, non-trivial challenges arise on the control and data paths when naively introducing modern high-performance disaggregation to these solutions.

We discuss these issues in depth in Section~\ref{sec:motivation} and identify key limitations, including low scalability, performance degradation, and high CPU utilization. 
%
These impact security, require unrealistic synchronizations that would congest the network, and can reduce overall throughput by up to 50\% while driving CPU usage above 80\%.
To address these challenges, we propose \projectname, a first-of-its-kind secure disaggregated storage system that extends the existing NVMe-oF capabilities to satisfy the CC threat model in a scalable and performant manner. 



We achieve this by introducing three key innovations summarized in Figure~\ref{fig:child_poster}.
First, we introduce a scalable control path.
Scalability is critical, as modern storage solutions consistently grow in size with SSD density doubling every 1.5 years~\cite{IntroducingLightningFlexible2016}.
Existing multi-SSD setups already reach 1 PB, and some vendors discuss developing single 1 PB drives~\cite{SamsungTalksPetabyte}.
A key observation that enables our control path is that disaggregating storage also opens a new opportunity to disaggregate the Merkle Tree (MT), which provides freshness~\cite{chakrabortiDmxProtectingVolumelevel2017, gueron:a_memory_encryption_engine_2016}.
As this increases the threat boundary, we introduce a novel control flow that distributes the necessary protections.
We perform MT management and verification on the storage node, saving communication overhead on the data path and addressing scalability issues of freshness.
%
%
\projectnameSpace also provides an IV leasing strategy for strict confidentiality on PB-sized storage.



Second, we optimize the data path by building on a key observation: many overheads in existing solutions can be eliminated. 
%
%
Instead of modifying the NVMe-oF protocol, which might require hardware changes, \projectnameSpace relies on request encapsulation to enable security on existing systems.
For that, we leverage a new design trend in high-performance SSDs: large block metadata as part of the logical block address (LBA)~\cite{280686}.
\projectnameSpace uses metadata to achieve crash-protected integrity and confidentiality at negligible cost, and to replace transport-layer solutions such as IPSec~\cite{ferguson1999cryptographic}, minimizing repeated encryptions on the critical path.
\projectnameSpace also introduces a novel tree organization for freshness: Hazel MT (\ourTree).
\ourTree addresses the shortcomings of a typical MT when applied to PB-scale drives by modifying its storage, verification, and update methods.
%
\ourTree supports crash protection, enables asynchronous writes, and no-overhead read-path verification. 
%

Third, we improve resource efficiency by leveraging the key observation that infrastructure tasks can be offloaded to network-specific accelerators such as smart NICs (sNICs), thereby alleviating CPU pressure~\cite{sriramanAccelerometerUnderstandingAcceleration2020, NVIDIABlueField3SNAP}. 
As this requires extending the trust boundary onto the sNIC, we exploit a new trend that equips them with CC capabilities~\cite{lalDataProcessingUnit2023}.
We use CC-capable sNICs as infrastructure security managers and as accelerators designed for network line rates. 
While we deploy \projectnameSpace on an sNIC to alleviate resource pressure and enable features such as GPU direct, \projectnameSpace can be leveraged as a CPU service.
It can also be deployed in a non-disaggregated setup, as our ideas are still applicable.

We prototype \projectnameSpace on an NVIDIA BlueField-3 using Storage Performance Development Kit (SPDK)~\cite{yangSPDKDevelopmentKit2017} and NVIDIA DOCA~\cite{nvidia_doca_2025}.
We comprehensively evaluate it across a range of synthetic patterns, benchmarks like IO500 and YCSB, and applications such as AI training. 
We show that \projectnameSpace can achieve as little as 1-2\% overheads.


\textbf{In summary, our contributions are}:
\begin{enumerate}
    \item Designing \projectname, a novel secure disaggregated storage system providing confidentiality, integrity, and freshness under the CC threat model, with its control path innovations.

    
    \item Introducing data path enhancements through NVMe metadata and a new MT type, \ourTree, offering optimized reads/writes. 
    
    
    \item Prototyping and open-sourcing\footnote{\url{https://github.com/spcl/hazel}} \projectnameSpace on top of SPDK with NVIDIA BlueField-3 offloading. 
    
    
    \item Evaluating \projectnameSpace on a real system, showing as little as 1-2\% overhead on synthetic benchmarks, IO500, YCSB, and a machine learning training pipeline.
\end{enumerate}

%% file: 3_motivation.tex
\section{Inadequacy of local secure storage}
\label{sec:motivation}

Linux's \textit{dm-crypt} and \textit{dm-integrity} are recommended solutions for TEEs~\cite{IntelTrustDomain}. 
\textit{dm-crypt} enables confidentiality by encrypting sectors.
Sectors are the smallest read and write units in NVMe SSDs.
They are typically either 512\,B or 4096\,B, and can have 0-8\,B of associated metadata.
%
%
Integrity is provided through \textit{dm-integrity} with cryptographic hashes.
In addition to these two solutions, \textit{dm-x}~\cite{chakrabortiDmxProtectingVolumelevel2017} can provide freshness by employing a Merkle Tree (MT).
Freshness protects against replay attacks where an adversary replaces a value with a stale one.
Such a temporal difference indirectly breaks integrity but is not detected by simple hashing.
%
%
%
Leaves of MTs are hashes of data computed for integrity protection. 
%
%
The branching factor of an MT defines how many children are hashed together to obtain parents.
Changes to the data trigger a hash recalculation at each level, which ultimately updates the root.
This guarantees online freshness, as changes to old values will manifest themselves with a different node value somewhere in the tree during subsequent read verification.
The root must be stored in a persistent, secure location to maintain offline freshness.

These kernel-level solutions are insufficient, as they were designed to secure \emph{local rather than disaggregated storage}. 
While providing confidentiality, integrity, and freshness, \textit{dm-crypt}, \textit{dm-integrity}, and \textit{dm-x} struggle when naively applied to protect disaggregated storage for three reasons: performance and resource usage, scalability, and security. 



%


%

\subsection{Scaling confidentiality and integrity}

%

Encryption schemes like AES~\cite{nistAES,dworkinRecommendationBlockCipher2010, dworkinRecommendationBlockCipher2007} provide confidentiality by dividing data into blocks and processing them with a secret key and a unique IV. 
%
Reusing IVs with the same key introduces collisions, weakening security.
For some ciphers, such as AES-XTS, this only leaks equality information.
However, for many authenticated encryption with associated data (AEAD) algorithms that provide confidentiality and integrity, such collisions represent severe security vulnerabilities.
For instance, collisions in AES-GCM are catastrophic, leaking plaintext data~\cite{kampanakis2024practical}.
As we show in Section~\ref{sec:resources}, AEAD algorithms achieve integrity considerably more efficiently than combining IV-resilient ciphers with separate hashing, making unique IVs essential.
National Institute of Standards and Technology (NIST) specifies the safe probability of collisions as $2^{-32}$~\cite{kampanakis2024practical}.

A common approach to implementing confidentiality in \textit{dm-crypt} is to allocate each storage partition a key and use the sector number as the IV.
Such an approach introduces collisions when writing to the same sector, since the key remains constant.
%
%
%
%
%
%
Using the sector number as an IV was attractive because it eliminated the need to store additional information per sector. 
However, when introducing integrity hashes, side storage becomes unavoidable, adding overhead (Section~\ref{sec:resources}). 
Two alternatives provide unique IVs on every write: random generation and counters. 
Once unique values are exhausted, a new encryption key must be provisioned.

\begin{table}[t]
\centering
\resizebox{\columnwidth}{!}{%
\begin{tabular}{@{}ccccc@{}}
\toprule
\multirow{2}{*}{Algorithm} & \multirow{2}{*}{\begin{tabular}[c]{@{}c@{}}Block size\\ {[}bits{]}\end{tabular}} & \multirow{2}{*}{\begin{tabular}[c]{@{}c@{}}IV size\\ {[}bits{]}\end{tabular}} & \multicolumn{2}{c}{Write capacity} \\ \cmidrule(l){4-5} 
                           &                                                                                  &                                                                               & Random     & Sequential            \\ \midrule
AES-XTS                    & 128                                                                              & 128                                                                           & 4.2 PB     & $5.4\cdot10^{24}$ PB  \\
AES-GCM                    & 128                                                                              & 96                                                                            & 64.1 GB    & $1.3\cdot10^{15}$ PB  \\
AEGIS128L                  & 256                                                                              & 128                                                                           & 8.4 PB     & $1.1\cdot10^{25}$ PB  \\
ChaCha20                   & 512                                                                              & 96                                                                            & 256.2 GB   & $5.1\cdot10^{15}$ PB  \\ \bottomrule
\end{tabular}%
}
\caption{Different encryption algorithms with the total amount of data that can be safely written using a single key. We use NIST guidelines for collision probabilities $p(n,d)$ and $n\approx\sqrt{-2\cdot2^{d}\log(1-p(n,d))}$, with constant cipher parameters: $n$, number of encryptions, $d$, number of IV bits.}
\label{tab:collision}
\vspace{-2.5em}
\end{table}

\paragraph{Random IVs:} Randomly sampled IVs are constrained by the birthday paradox~\cite{suzuki2006birthday}, which limits the number of writes per key before collisions.
Table~\ref{tab:collision} shows maximum safe write volumes for a single key.
In the best case, random IVs allow up to 8.4\,PB of writes using the same key. 
While this suffices for a local SSD, disaggregated PB-scale storage requires considerably more.
Since an SSD's average lifetime is 10,000 write cycles, the required write capacity is in the thousands of PBs.
This limitation necessitates key rotation.
Existing approaches do not address questions such as how to synchronize write counts of a given key across multiple nodes (tenant instances), how to record which blocks were written with which keys, how to generate the keys and synchronize them between nodes, and where to persistently store them in a CC compliant manner.

\begin{figure*}[b] 
    \centering
    \includegraphics[width=\textwidth]{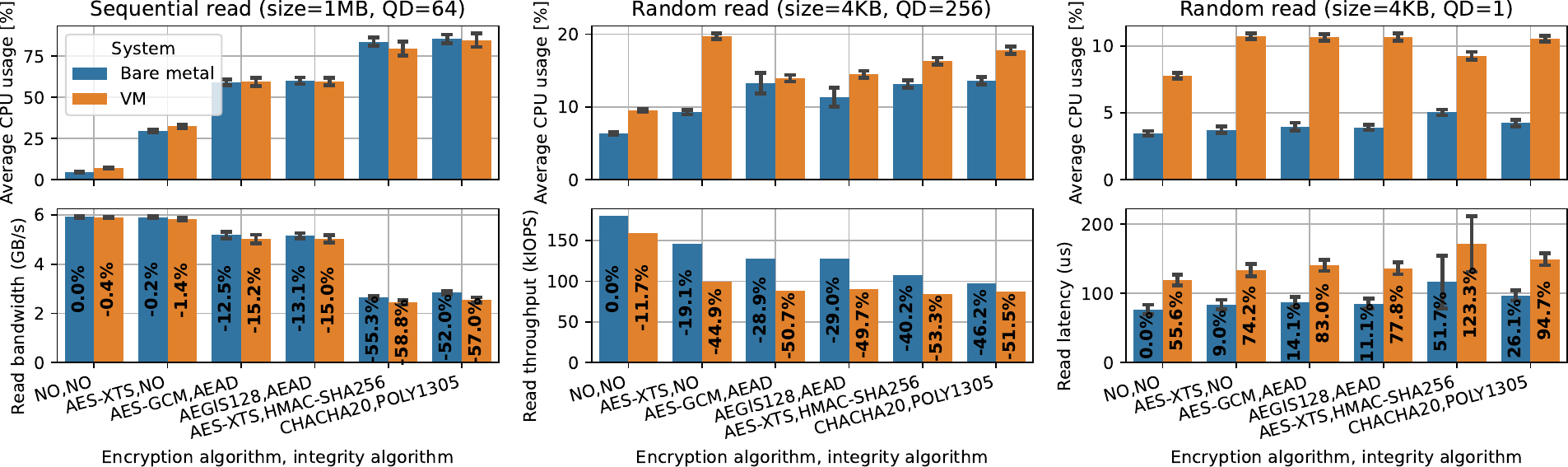} 
    \caption{Read throughput, IOPS, and latency across different encryption and integrity algorithms with average CPU usage on a hyperthreaded system (50\% CPU usage involves all physical cores). Overheads are relative to unprotected bare metal.}
    \label{fig:overheads}
\end{figure*}


\paragraph{Counter IVs:} Increasing a counter during each write provides IVs not bound by the birthday paradox.
While almost limitless, the default counter IVs in \textit{dm-crypt} are not scalable. 
Multiple instances must coordinate to avoid overlapping counters, requiring synchronization across instances.
Such coordination is manageable at a small scale but quickly becomes infeasible in a cluster setting, where all-to-all synchronization would severely slow the data path and require instances to trust one another.

%
\paragraph{\projectname's solution:} Therefore, in disaggregated environments, where users are typically provided access to a local, small SSD and a remote, large just-a-bunch-of-flash (JBOF) array, scalability demands a separate control path and infrastructure not found in existing systems.
%
%
%
We design a scalable counter-leasing protocol combined with a key management framework for derivation and storage. 
%
This enables multiple instances to use disaggregated JBOFs with per-partition counter allocation without incurring synchronization performance overheads.
%
%

\subsection{Overheads of confidentiality and integrity}
\label{sec:resources}
Another issue with these solutions is their resource and performance costs. 
Figure~\ref{fig:overheads} shows throughput, input-output operations (IOPS), latency, and CPU usage for sequential and random reads of varying sizes and queue depths (QD), across different encryption (\textit{dm-crypt}) and integrity (\textit{dm-integrity}) algorithms. 
%

While AES-XTS confidentiality can provide raw disk performance, even in our single-SSD setup, it consumes about 30\% CPU.
Adding integrity increases overheads to 50\% for AEAD algorithms and 80\% for algorithms that compute integrity and encryption in sequence.
As the CPU becomes saturated, the performance also suffers, especially with combined algorithms, where the overheads can reach up to 50\%.
%
Such a high CPU usage dedicated to security operations is expensive and prevents efficient parallel data processing.
These overheads would be exacerbated in JBOFs, which can deliver several times the throughput of our testbed.

Overheads arise not only from excessive resource usage. 
Synchronous CPU accelerators (e.g., AESNI~\cite{hofemeier2012introduction}) block parallel request processing, kernel context switches introduce latency, and integrity hashes incur additional I/O. 
%
%
%
Specifically, \textit{dm-integrity} places integrity hashes and non-constant IVs in separate metadata sectors that require additional storage.
Thus, reads require an additional read, and writes require an additional read-modify-write.
While \textit{dm-crypt} in the 6.14 kernel release introduces support for NVMe metadata, this only stores hashes and not the IVs.
Such a design reduces IOPS, increases latency, and necessitates journaling to enable crash protection.
As journaling requires writing the data also to the journal, write performance can be reduced by up to 50\%.

\paragraph{\projectname's solution:} We address these bottlenecks by leveraging NVMe metadata to its full potential: storing not only IVs and hashes but also additional security information.
%
To eliminate CPU overhead and improve performance, we move operations out of the host kernel (where all three \textit{dm-x}, \textit{dm-crypt}, and \textit{dm-integrity} reside) into user space.
We also offload operations to modern sNICs, which are on the critical data path in either case.
%
%
Unlike synchronous x86 CPUs, sNICs provide asynchronous cryptographic accelerators designed for line-rate processing at lower power~\cite{guoLEEDLowPowerFast2023, xingPortableEndtoendNetwork2022}, enabling secure operations to run in parallel with IO requests.


\subsection{Scaling and overheads of freshness}

A significant challenge with freshness is ensuring it at scale.
\textit{dm-x}, designed to run on an instance, requires global consistency across all instances sharing a block device.
If multiple instances issue writes concurrently, they must coordinate to maintain a coherent MT.
For example, if one instance writes sector zero while another writes sector one, both must synchronize to update the parent without conflicts.
Such coordination scales poorly, as it requires mutual trust between instances and frequent cross-instance communication to share secure state and the persistently stored root.
This adds network overhead and slows the critical path.

Furthermore, the above discussion on overheads does not account for freshness.
\textit{dm-x} also uses journaling for freshness-crash protection.
On top, it places the MT on the disk, which implies that updating and verifying it might require many additional reads or writes, all of which are on the critical path.
\textit{dm-x} with \textit{dm-crypt} and \textit{dm-integrity} have been reported to introduce 54\% and 63\% throughput degradation of synthetic sequential reads and writes, respectively~\cite {chakrabortiDmxProtectingVolumelevel2017}.
Additionally, this has been measured on weak SSDs with a maximum throughput of 280\,MB/s for reads and 240\,MB/s for writes, rather than a GB/s level of modern storage.

\paragraph{\projectnameSpace's solution:} We address this by disaggregating the integrity tree to the JBOF and outlining a protocol that ensures trust and supports multiple users. 
We further propose a novel \ourTree (Section~\ref{sec:tree}), that modifies typical verification and update algorithms to be multithreaded, cached, and eventually consistent.
It also introduces its own metadata that enables a zero-overhead freshness verification.
These create fast read and write paths without synchronization, minimizing the performance overheads.


%% file: 4_design.tex
\section{Design of \projectnameSpace}

Our design focuses on two principles: performance and scalability, while guaranteeing critical security properties of confidentiality, integrity, and freshness.
We present our solution in the most generalizable form possible.
This allows us to reach the broadest audience while maintaining flexibility for practitioners.
For example, if the network is freshness protected (e.g., Ultra Ethernet encryption~\cite{hoeflerUltraEthernetsDesign2025}), \projectname's network freshness header is surplus, allowing for more IV and key ID bits.
We first outline our threat model (Section~\ref{sec:threat_model}), and follow with control (Section~\ref{sec:control_path}) and data paths (Section~\ref{sec:data_path}).

\subsection{Storage and threat model}
\label{sec:threat_model}

We assume SSDs are disaggregated via NVMe-oF, an extension of the NVMe protocol that enables high-speed, low-latency data transfer between hosts and storage devices across the network~\cite{NVMeFabricsSpecification2021}.
While our design is based on NVMe-oF, its principles can apply to other secure storage systems.
We adopt a standard CC threat model~\cite{mulliganConfidentialComputingBrave2021} with both privileged adversaries (e.g., a cluster administrator) and unprivileged ones (e.g., other users) targeting NVMe-oF and the storage devices.
Adversaries might be local (on participating nodes) or remote (on the network), and can observe, replay, or tamper with control- and data-plane requests and responses.
Recently, NVMe-oF has incorporated several security mechanisms, such as TLS or IPSec transport protection.
However, securing the transport is not a silver bullet, as in CC, the data stored at rest must also be protected.
Thus, \projectname's primary goal is to protect data at rest on untrusted, disaggregated storage devices serving TEEs, ensuring confidentiality, integrity, and freshness.
Following prior work such as dm-x~\cite{chakrabortiDmxProtectingVolumelevel2017}, we do not trust the storage devices themselves.

We assume adversaries that may attempt replay or modification attacks, but cannot compromise TEE execution through side channel or transient-execution attacks~\cite{feiSecurityVulnerabilitiesSGX2021}.
While these may expose TEE information, they are orthogonal to our work.
We also do not focus on denial-of-service attacks, which would violate the service guarantees and reveal privileged attackers.
All \projectname\ services are assumed to run inside attested, vulnerability-free TEEs, with access to a small trusted non-volatile memory for storing persistent security-sensitive state (see Section~\ref{sec:write_path}).



\subsection{Control path}
\label{sec:control_path}

\begin{figure*}[t] 
    \centering
    \includegraphics[width=\textwidth]{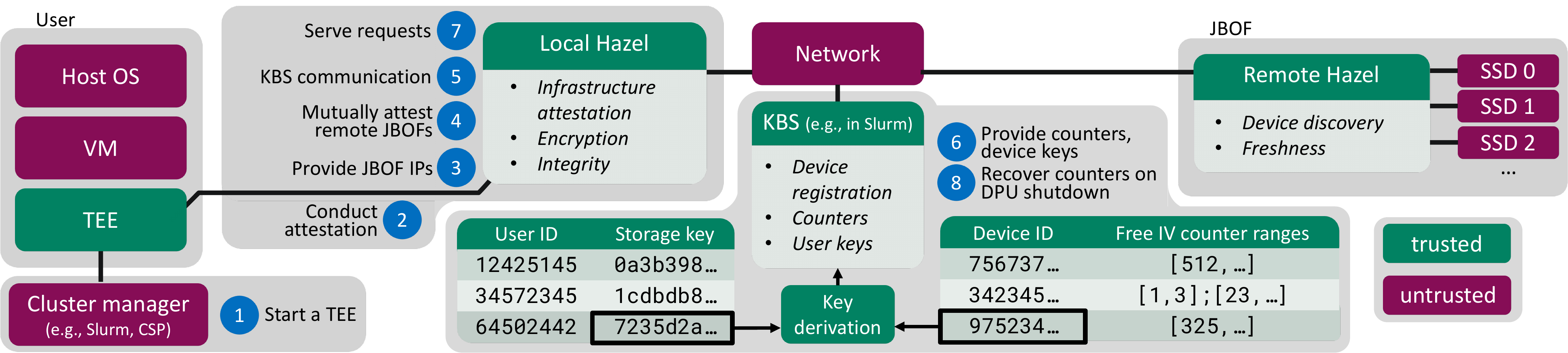} 
    \caption{\projectnameSpace control path (Section~\ref{sec:control_path}) starts with the cluster manager scheduling a TEE on the corresponding host~\protect\negcircnum{1}.
%
%
Afterwards, TEE attests the local \projectnameSpace instance~\protect\negcircnum{3} and provides JBOF IPs~\protect\negcircnum{4}. 
The local \projectnameSpace uses these to open connections through mutual attestation~\cite{menetreyAttestationMechanismsTrusted2022, chenMAGEMutualAttestation2022}~\protect\negcircnum{5}.
The local \projectnameSpace service also provides the user identity to KBS~\protect\negcircnum{6}, leveraging the connection opened at boot and kept alive.
KBS yields security details necessary for starting operation~\protect\negcircnum{7} and handles shutdown~\protect\negcircnum{8}.
Steps \protect\negcircnum{4} and \protect\negcircnum{5} are independent from \protect\negcircnum{3} and \protect\negcircnum{6}, and can be conducted in parallel.
}
    \label{fig:control_path}
\end{figure*}

Figure~\ref{fig:control_path} shows an overview of \projectname's control path with untrusted (burgundy) and trusted (green) entities.
The latter are brought into the trusted computing base (TCB) through attestation. 
We differentiate two types of \projectname\ instances: local and remote.
These are used for different purposes, as explained in the data path description.
We split the control path into three main phases: establishing trust, initializing security, and operation.
%

\subsubsection{Establishing trust\,\protect\negcircnum{1}-\,\protect\negcircnum{5}} 

%
%
%
A central idea we introduce at this step is that the local instance also functions as a trust gate for the infrastructure.
It not only secures its environment but also extends trust to the rest of the storage network, relieving the user from this responsibility.
%
%
%
The local \projectnameSpace service enforces network security and establishes trust with remote \projectname{} instances and the Key Broker Service (KBS). 
For that, the KBS and \projectnameSpace conduct mutual attestation~\cite{menetreyAttestationMechanismsTrusted2022, chenMAGEMutualAttestation2022}, as they need to ensure that both sides are in the correct state. 
KBS can be incorporated into a cluster manager (e.g., Slurm~\cite{yooSLURMSimpleLinux2003}) or be a standalone service (e.g., CSP-based). 
%
%
%

Such an approach has performance advantages over the user attesting each agent in the network. 
As the \projectnameSpace service handles these remote attestations, it can perform them at boot time and keep the already-opened secure connections (e.g., TLS) alive.
The user does not need to additionally attest the KBS, and potentially tens to hundreds of devices, each taking at least several milliseconds~\cite{misonoConfidentialVMsExplained2024}. 
%
%
Connections to JBOFs can also be opened preemptively if the CSP or cluster manager knows it will allocate specific hosts to specific JBOFs. 
This minimizes the initialization cost of \projectname, which is crucial for short-lived workloads affected by startup overheads, such as Function-as-a-Service (FaaS)~\cite{copikSeBSServerlessBenchmark2021}. 

\subsubsection{Initializing security\,\protect\negcircnum{6}}
\label{sec:initializing_security}

KBS manages counters and keys.

\paragraph{Counter leasing:} \projectnameSpace obtains unique IVs through a counter-leasing protocol.
When a remote \projectnameSpace instance discovers a new SSD device, the KBS starts a $b$ bit range by creating a list per device: \verb|[(0, |$2^b$\verb|)]|. 
Any \projectnameSpace service that would like to access a storage device sends a request to the KBS with the appropriate device ID.
The KBS then leases a range of counters corresponding to 1\,TB, large enough to avoid frequent requests for new ones.
Using a counter allows \projectnameSpace to allocate an astronomically large number of ranges (e.g., $1.3\cdot10^{18}$ for AES-GCM), which can span the SSD's lifetime without refreshing keys.
The space required for such a counter list scales linearly with the number of SSD devices rather than the number of user partitions. 
Local \projectnameSpace service caches these counters as long as possible, even after the shutdown of the associated TEE. 
This ensures multiple user jobs (tenant instances) can be run on the same host without repeatedly leasing and returning the same ranges.
If a range needs to be evicted, \projectnameSpace sends it back to the KBS for reuse, even if partially used.
The KBS adds it to its per-device list, ensuring compaction, and always returns 1\,TB of write capacity, which might mean smaller subranges are provided.
%
%
%

\begin{figure*}[t] 
    \centering
    \includegraphics[width=\textwidth]{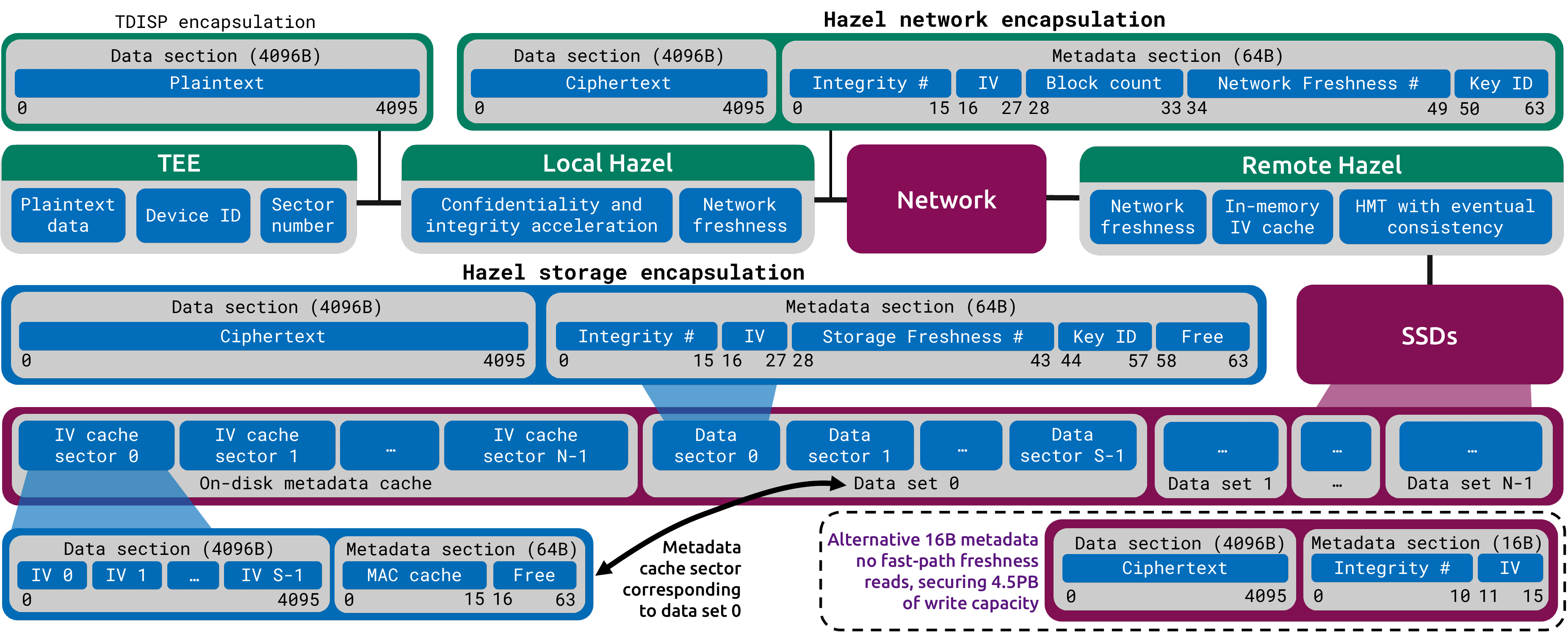} 
    \caption{\projectnameSpace's data path, including the security encapsulation, and its SSD sector layout zoom-in.}
    \label{fig:sector_layout}
\end{figure*}

\paragraph{Key management:} \projectnameSpace also manages keys for confidentiality and integrity guarantees.
%
Instead of storing user-partition keys, we base our management on key derivation.
A user-device key $k_d$ is derived from a cryptographic hash function $H$ for a device with ID $i$, and a user storage key $k_s$ using an HMAC~\cite{krawczykHMACKeyedHashingMessage1997}
$
k_d = \text{HMAC}_{k_s}(i) = H\left((k_s\oplus opad)||H\left((k_s\oplus ipad)||i\right)\right)   
$
with $\oplus$ representing the XOR operation, $||$ concatenation, and $opad$, $ipad$ outer and inner padding (repeated \verb|0x5c| and \verb|0x36| correspondingly). 
Unique identifiers such as SSD EUI64 can be used as device ID $i$.

We placed key derivation on the KBS rather than the local \projectnameSpace instance, as this limits potential exposure of the master user key in case of a local data leak.
Such isolation is preferable to local solutions such as \textit{dm-crypt}, which would require a global user storage key.
%
%
We also decided to operate at the device level rather than NVMe controllers, namespaces, or other partitions, as any change to these (e.g., switching controllers) would require additional tracking and potentially full-disk re-encryption. 
%


\subsubsection{Operation\,\protect\negcircnum{7}-\,\protect\negcircnum{8}} 

\projectnameSpace then moves to the data plane, serving requests. On shutdown, it returns the counter ranges to KBS, erases local information, and closes connections. 
\vspace{-0.75em}


\subsection{Data path}
\label{sec:data_path}
\projectnameSpace minimizes the overheads on the critical path.
We leverage a key concept of NVMe metadata to encapsulate secure information, without modifying the NVMe-oF protocol. 
Figure~\ref{fig:sector_layout} shows the encapsulations that \projectnameSpace introduces within this 64\,B protocol metadata, including data fields and sector layouts.
%
%


\label{sec:sector_layout}


Each SSD is split into two distinct sections: metadata and data.
Data sectors hold ciphertexts, and their metadata contains hashes, IVs, key identifiers, and freshness caching information used for the fast read path.
We group $S$ data sectors into \textit{data sets}.
Each data set has a corresponding metadata sector placed at the beginning of the SSD.
Metadata sectors hold aggregated IVs of data sectors and serve as an on-disk cache to optimize freshness.
%
%
%
We calibrate $S$ such that as many IVs as possible fit in one metadata sector.
For a typical IV of 12B and a sector size of 4096B, $S=340$, yielding \projectname's footprint on storage of only 0.29\%. 
%


\subsubsection{Local \projectnameSpace write path}
\label{sec:write_path}
Writes start at the TEE as a request consisting of data and NVMe-oF remote device address.
The request is then transported to the local \projectname{} service, which encrypts and integrity protects it.
If the local \projectnameSpace is offloaded, this might include secure protocol such as TDISP~\cite{tdisp2022} (Section~\ref{sec:related_work}).

\paragraph{Confidentiality and integrity:} Combining integrity and encryption algorithms one after the other adds considerable overhead (Section~\ref{sec:motivation}).
Thus, leveraging an AEAD algorithm $E$, \projectnameSpace obtains ciphertext $C$ and integrity hash $H_i$ as: $C,H_i = E(k,$\text{sector number}, \text{IV}, \text{write data})
where $k$ is a key derived from the KBS provided user-device key $k_d$ as 
$
k = HMAC_{k_d}(\text{key ID bits})
$, and sector number forms AD.
Including the sector number ensures that sectors cannot be swapped without detection.
%
The device-specific key ensures that writes to the same sector with the same IV differ across disks. 
After each encryption, the IV is increased by the number of encryption blocks. 
While not strictly necessary from a security perspective, including the key ID makes key rotation easier.

\paragraph{Network freshness:} Disaggregating the freshness protections to the remote side is crucial to enable scalability.
However, this necessitates protecting sent sectors as \projectnameSpace operates in a CC environment.
Although intuitive, IPsec has considerable drawbacks in our setting.
It adds latency to the critical path and decreases the throughput. 
Furthermore, IPsec’s per-IP security model clashes with NVMe multi-service deployments~\cite{taranovNeVerMoreExploitingRDMA2022a}. 
Instead of using IPSec, \projectnameSpace leverages already applied protections.
Confidentiality and integrity are already provided by local \projectnameSpace, and protecting headers adds little value besides out-of-scope DoS attacks. 
%
%
Freshness needs also to be guaranteed over the network, for which we use windowing~\cite{ferguson1999cryptographic}. 

Once a secure channel is opened between the local \projectnameSpace and the remote JBOF, the \projectnameSpace instances agree on a communication key $k$ and a pair of random start counters, one for sending and one for receiving. 
Writes use the former to obtain a block count $j$ and compute a network freshness hash $H_n$ per sector of each request as $H_n=\text{HMAC}_k(IV||j)$, 
with $||$ representing concatenation. 
$H$ is then placed into the corresponding sector's metadata.
As $j$ is 48 bits, $k$ needs rolling at most every 1.4 years, assuming a continuous 200\,Gbit/s usage.
Hashing only the IV does not add substantial overhead and is secure, as it is cryptographically bound with the data through $H_i$.
%
%
Through this we avoid repeated encryption and integrity protection at the cost of increased metadata usage.
%
The network freshness is then verified at the JBOF using a sliding window: writes with a counter at most $T$ lower than the current maximum are accepted.
$T$ enables asynchronous request delivery. 

\begin{figure*}[t]
\centering

\begin{minipage}[t]{0.36\textwidth}\vspace{0pt}
  \centering
  \includegraphics[height=\panelH,keepaspectratio]{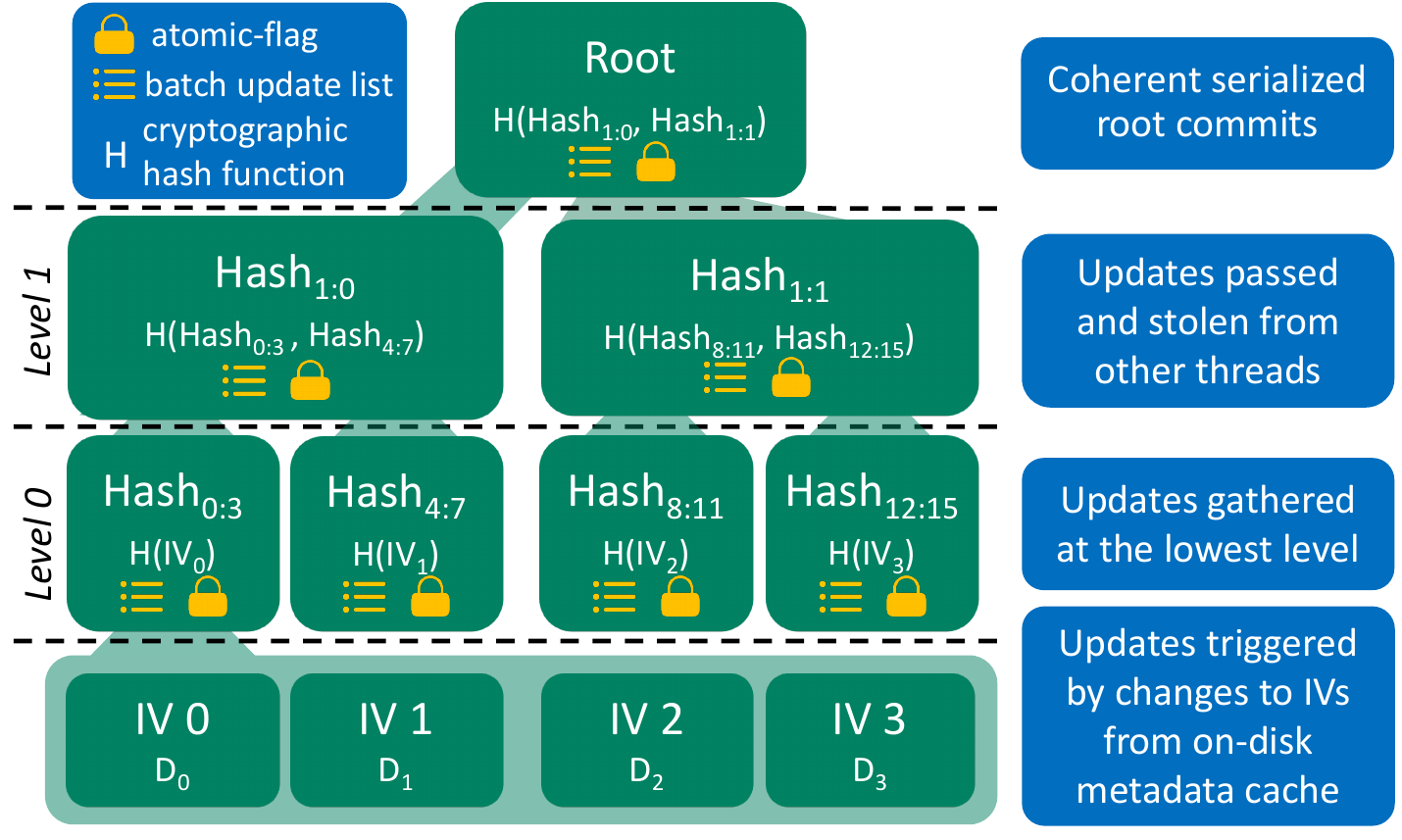}
  \captionof{figure}{The overview of HMT design and the update strategy involving batch lists and serialized root commits for consistency.}
  \label{fig:tree_description}
\end{minipage}\hfill
\begin{minipage}[t]{0.30\textwidth}\vspace{0pt}
  \centering
  \includegraphics[height=\panelH,keepaspectratio]{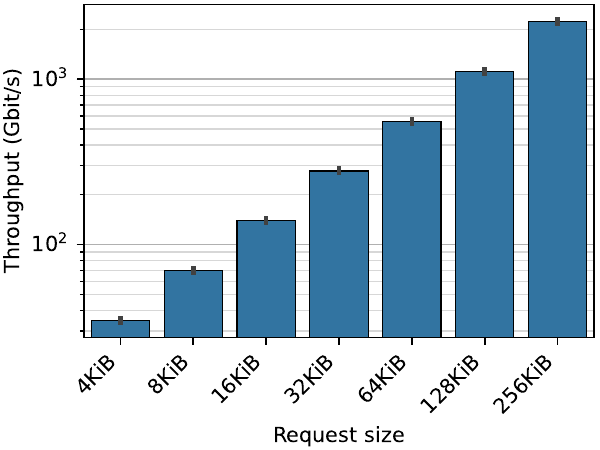}
  \captionof{figure}{HMT throughput of processing updates against the number of batched requests.}
  \label{fig:tree_tput}
\end{minipage}\hfill
\begin{minipage}[t]{0.30\textwidth}\vspace{0pt}
  \centering
  \includegraphics[height=\panelH,keepaspectratio]{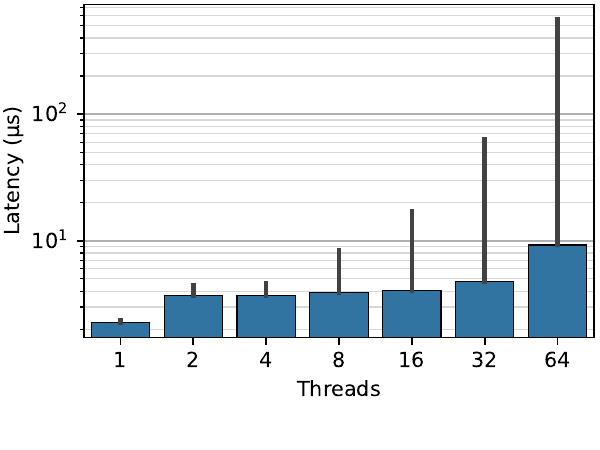}
  \captionof{figure}{HMT median and P95 latency against the number of hashing threads.}
  \label{fig:tree_latency}
\end{minipage}
\end{figure*}

\subsubsection{Remote \projectnameSpace write path:}
\label{sec:tree}
The data must then be committed to the drives and the freshness tree on the remote side.
\projectnameSpace introduces the Hazel Merkle Tree (\ourTree).
Similar to a Bonsai MT~\cite{rogersUsingAddressIndependent2007}, \ourTree uses IVs instead of data as its leaves, reducing MT's compute and memory footprint.
However, it considerably differs in \emph{storage}, \emph{verification}, and \emph{update} methods. 
The key challenge for \ourTree is to maintain consistency between the in-memory IV cache and the disks while ensuring multithreaded performance.
\ourTree leverages two novel concepts: LBA metadata for fast reads, and eventual consistency and batching for fast writes.

\paragraph{Storage:} Apart from its root and a small block cache, an MT in \textit{dm-x} is stored entirely on disk.
Loading parts of the tree introduces non-deterministic latency for both reads and writes, reducing IOPS and throughput.
On the other hand, storing the whole tree in memory is expensive.
%
%
Given an SSD with $N$ sectors and $B$ bytes per hash, such a tree is at least $B(N+1)$ bytes (branching factor $N$ and a single root node).
For a hash of $B$=16~B and a 1~PB drive ($N$=$244\cdot10^9$), the tree is 3.9~TB. 
Equipping JBOFs with that much memory is expensive.
 
To address this \ourTree is created leveraging two branching factors: one for the leaf data $D_d$, and one for the tree $D_t$.
We store the tree in memory without the underlying zeroth level, which accounts for most of its size.
This allows \ourTree to scale considerably better as the in-memory part of the tree is only $B(N+1)/D_d$. 
As mentioned in Section~\ref{sec:sector_layout}, we batch $S$ IVs in a single metadata sector on the SSD.
We set $D_d=S$, thereby making the tree size manageable.
For example, for 1 PB, $S=340$, and representative and minimum branching factors of 16 and 2, the tree sizes are 12.3~GB and 23~GB, respectively, which can fit even on a relatively small DRAM memory of an sNIC.
This optimization is crucial for performance, as it allows us to always maintain the tree without data leaves in memory, enabling the zero-overhead read path (Section~\ref{sec:read_path}). 
\vspace{-0.5em}
\paragraph{Updates:} Each node in an \ourTree contains a hash, a lock, and a pointer to a batch update list, as shown in Figure~\ref{fig:tree_description}. 
Each write enqueues an update to the corresponding leaf.
Hashing threads (hashers) pick up these task lists, hash the children, and modify parents by concatenating the new updates.
Such a design has two advantages.
First, it enables crash protection through a coherent view of the children, which we discuss later in the section.
Second, it allows request batching at the lowest and intermediate levels.

We implemented an \ourTree on a 1 PB disk to measure its hashing throughput by timing the processing of 10 M updates.
After conducting a sweep, we selected five hashing threads that performed best.
This number can vary considerably between machines, as we observed on another setup.
We also used a branching factor of 16 and implemented contention-heavy repeated writes to the same location.
%
%
Figure~\ref{fig:tree_tput} shows how the effective throughput of storage scales with the size of each write.
Note that for smaller writes, the tree's throughput can be a bottleneck.
Batching enables us to combine these requests, thereby increasing their effective size.

We also measured the time required to complete tasks. 
Figure~\ref{fig:tree_latency} shows the median and 95th percentile of the latency.
We observe that as we increase the number of hashers required for throughput, the 95th percentile grows rapidly.
Because the tree, even for 1 PB, is shallow, containing only nine levels, the contention is high.
Exposing this contention on the write path is expensive.
For this reason, \ourTree introduces eventual consistency (EC).

\textit{dm-x} handles writes by updating the tree and then writing the data to the disk. 
This directly puts hashing and, for \textit{dm-x}, potentially reading tree blocks from the SSD on the critical path.
%
%
In \ourTree, we place hashing out of the critical path by completing write requests even before the tree update is done. 
%
%
%
%
This makes the tree eventually consistent with the disk rather than continuously tracking its state.

\paragraph{Coherency and crash recovery:} While it resolves the latency issue, EC also needs to ensure a coherent state between disks and the tree.
In the case of parallel writes to the same location, the final state on the disk is nondeterministic as writes and their completions are asynchronous.
While this is fine in a typical setting, in our case, it leads the tree to lose coherence with the disk.
Thus, we allow only one writer per sector, synchronizing writes and ensuring that the disk order matches the tree update order.
%

Coherence is also a necessary ingredient to ensure crash recovery. 
While integrity is easy to provide because writes at the sector level are guaranteed to be atomic under the NVMe standard for both data and metadata, such protection is not trivial for freshness.
This is because the root of any MT is not atomically updated during disk writes.
The strategy provided by \textit{dm-integrity} or \textit{dm-x} is to use journaling, which involves first reading and saving the old data, then writing the new data.
However, this approach introduces overhead for IOPS, throughput, and storage. 
%
%
We resolve this by keeping track of a per-sector tuple of \verb|[status, location, old IV, new IV]| for each write request.
Using the new and old IVs, we can ensure the tree's state can be recovered in case of a crash.
However, these write requests must be stored in persistently crash-protected memory, alongside the tree's root.
This can be achieved by leveraging non-volatile SRAM, which involves using a standard IP module~\cite{NvSRAMNonvolatileSRAM, ComparisonBatteryBacked}.
As each request requires 24B, assuming a 200~Gbit/s network with each hashing update taking up to 40 us, the buffer for all requests requires only 3.5~KB, costing 1.75 cents, assuming the upper bound for SRAM price at \$5000/GB~\cite{lanzaGrowingMemristorIndustry2025}. 

\paragraph{IV cache design:} The IV cache enables reuse for sequential workloads.
The hashers need to know the IVs of all $S$ leaves within the same data set to conduct an update.
To avoid reading hundreds of sectors and their IVs stored in each sector's metadata, we introduced a metadata cache at the beginning of each SSD.
We additionally use an in-memory, write-back cache based on these sectors.
When a write occurs to a given data set, we fetch the corresponding metadata sector into the IV cache and update it, keeping it always up-to-date with the disk's state.
These caches could be eliminated if the NVMe specification allowed reading only the metadata portion of a sector rather than the whole sector.
%
%



\subsubsection{Read path}
\label{sec:read_path}
%
The request is first defined in the TEE and forwarded to the remote JBOF, where it is read from the disk. 
Once completed, freshness checks are conducted.

\paragraph{Read fast path:} \projectnameSpace introduces a novel freshness fast path.
During writes, alongside the data, we also store in the sector's metadata a secure authentication hash binding the IV and the parent's hash from \ourTree.
The \emph{verification} of \ourTree first compares the parent of the given node with the authenticated hash in the sector metadata.
If these fit, the freshness checks conclude.
If not, a complete freshness check is required. 
It involves fetching the metadata block into the in-memory IV cache and comparing the IVs with those read from disk. 
Freshness misses occur when a prior write updates the parent hash, making the sector metadata freshness cache stale. 
To avoid it, one can conduct only large writes or periodically run a job to ensure the metadata cache in each block is up to date (like thrashing in HDDs). 
%
%

%% file: 5_implementation.tex
\section{\projectname's implementation}
\label{sec:implementation}

We prototyped \projectnameSpace on top of SPDK~\cite{yangSPDKDevelopmentKit2017}. 
SPDK enabled us to conveniently expose \projectnameSpace instances in user space as virtual block devices (vbdevs) while ensuring maximum performance.
Our implementation has around 6K lines of C code that we release.
We used AES-GCM for encryption and BLAKE3~\cite{aumassonBLAKE2SimplerSmaller2013} for hashing.

\paragraph{Local instance: } We implemented the local \projectnameSpace service on top of an NVIDIA BlueField-3 Data Processing Unit (DPU)~\cite{bursteinNvidiaDataCenter2021}.
To access DPU's cryptographic Generic Global Accelerators (GGA), we used NVIDIA Data-Center-on-a-Chip Architecture (DOCA)~SDK~\cite{nvidia_doca_2025}.

We addressed some of DOCA's limitations and modified SPDK to work efficiently with it. 
Write requests arrive in the local \projectnameSpace vbdev as \verb|iovec| buffers with no metadata fields.
This data needs to be provided to the GGA accelerators as DOCA tasks.
However, any buffers provided to DOCA must first be registered with its \verb|mmap| function.
Putting this on the critical path is not viable as it takes on the order of milliseconds to complete.
Thus, we modified the SPDK buffer pool to register the SPDK \verb|malloc|'ed memory with DOCA a priori.
We then exposed this mapping globally so that Hazel vbdev can access it a couple of layers higher in the stack.
Our approach applies to future applications using both SPDK and DOCA buffers.

While sharing source and destination buffers is typically optimal, in DOCA, these two must be different, which requires the introduction of temporary buffers allocated from a pre-registered buffer pool.
We found that temporarily copying incoming requests to separate source buffers can bring more than 20\% performance overhead as the DPU cores are relatively weak.

Such data movement costs are also the main reason we avoided using associated data (AD), as it requires copying the data to a temporary buffer and placing AD before the source buffer to conform to the DOCA format.
%
To avoid this, we split the encryption algorithm's underlying IV between the sector number and the actual IV.
For our AES-GCM, the IV is 96 bits long, where we allocate the top 38 bits for sector numbers (1.13~PB of storage) and the bottom 58 bits for the IV (1.18$\cdot10^6$~PB of write capacity). 


DOCA cryptographic GGAs support only contiguous buffers that result in a single hash.
Thus, we define tasks per storage sector.
We batch these tasks and their completions together to reduce GGA MMIO overhead.
%
We also waited until all corresponding single-sector tasks were complete for large reads/writes, to be able to forward a single read/write IO to the remote \projectname.
We found this to be more performant than splitting into single-sector IOs.




\paragraph{Remote instance:} We implement an asynchronous least recently used (LRU) IV cache using a combination of a hash map and a doubly-linked list. 
This combination enables O(1) lookups (hash map) and LRU insertions (double-linked list) while not considerably increasing the memory requirements of the cache.
Reads or writes accessing a given block move it to the beginning of the LRU list.
%

If a cache entry is not present, it can be scheduled for reading. 
After completion, callbacks for each requester would be run on their corresponding SPDK threads.
%
%
If a new cache entry is needed, the last entry in the LRU list is written back, and the data is replaced with the requested sector. 
The new cache block is hashed and verified with the in-memory \ourTree to ensure freshness.

We implement \ourTree as arrays of hashes, \verb|atomic_bool| locks, and pointers to update lists.
We chose a branching factor of 16, as it optimized the tree's height while still allowing all 16 children to be hashed fast.
%
During writing, we use eventual consistency with four POSIX threads that operate similarly to an asynchronous accelerator. 
Tasks are scheduled in a round-robin manner among the threads traversing the tree until the root is reached.
We implement implicit ordering using the lock array index to avoid deadlocks.

%% file: 6_evaluation.tex
\section{\projectname's evaluation}
\label{sec:evaluation}
We address the following four research questions:
\begin{enumerate}
    \item[Q1] What are \projectnameSpace's overheads for synthetic patterns?
    \item[Q2] How does the performance of the freshness read path change as the metadata cache gets polluted?
    \item[Q3] What is the influence of eventual consistency on writes?
    \item[Q4] What is the impact of \projectnameSpace on benchmarks and applications?
\end{enumerate}

\begin{figure*}[] 
    \centering
    \includegraphics[width=\textwidth]{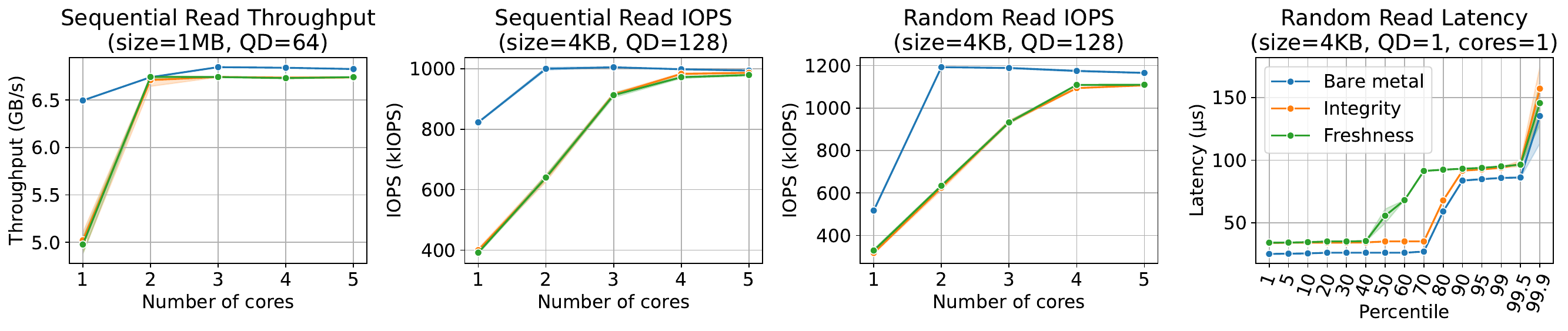} 
    \caption{Performance of \projectnameSpace across different read patterns.}
    \label{fig:combined_read}
    \vspace{-1em}
\end{figure*}

\vspace{-1em}
\subsection{Experimental setup}
\label{sec:setup}

We establish a two-host client-server setup: a user host and a storage server.
The former is a single-socket, 16-core, SMT-enabled AMD EPYC 7313P with 2x16 GB 3200 MT/s DDR4 memory running Ubuntu 22.04.5 LTS and Linux 5.15.
The latter is a double-socket, 40-core, SMT-enabled Intel Xeon Platinum 8460Y+ with 8x16 GB 4800 MT/s DDR5 memory, running Ubuntu 22.04.5 LTS and Linux 5.15.
The user has a 1.6 TB Samsung MZPLL1T6HEHP configured with 4 KiB sectors and no metadata, while the storage server has a 3.84 TB Western Digital DC SN655 configured with 4 KiB sectors and 64 B of metadata.
For the former, we used a 100 GB \verb|LVM| partition to measure performance.
Both are equipped with an NVIDIA BlueField-3 (BF3) NIC and are connected back-to-back with a 100~Gbit/s Ethernet link.
The user BF3 is configured as a DPU, while the storage BF3 is configured as a NIC.
The DPU is a 16-core ARM Cortex-A78AE with 32 GB 5200 MT/s DDR5 memory running Ubuntu 22.04.5 LTS and Linux 5.15.
The user and its BF3 use MOFED 24.10 and DOCA 2.9, while the storage server and its BF3 use MOFED 24.07 and DOCA 2.8.
We used GCC version 11.4, SPDK version 23.01 commit \verb|34edd9f1b|, BLAKE3 version 1.8.2 commit \verb|3a90f0f|.
For each synthetic experiment, we collected 10 s of measurements over at least 10 iterations.
For the benchmarks and applications, we ran each experiment at least three times.

\subsection{Read performance (Q1)}

\begin{figure*}[b] 
    \centering
    \includegraphics[width=\textwidth]{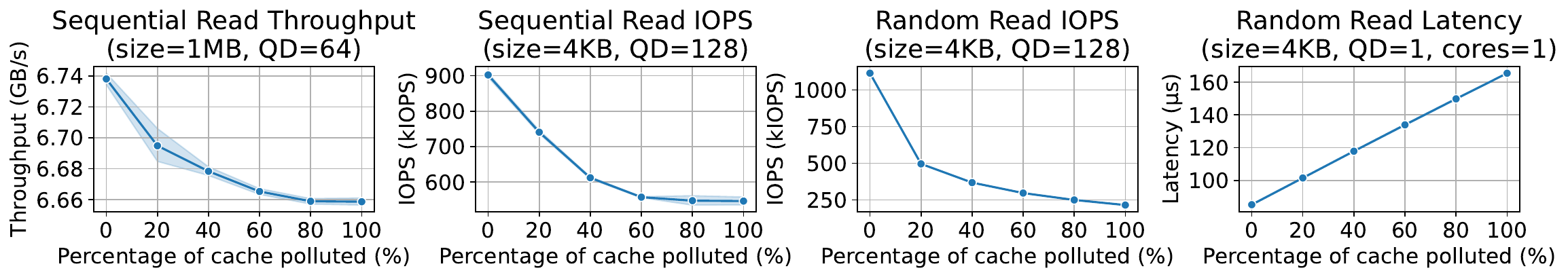} 
    \caption{Scaling of freshness performance as we modify the proportion of the metadata cache that is polluted.}
    \label{fig:cache_pollution}
    \vspace{-1em}
\end{figure*}


We evaluated the performance of \projectnameSpace using \verb|fio|'s SPDK backend when running four read patterns representing different types of workloads.
Bare metal is a deployment without any security, used as a performance baseline.
Integrity is using only the local \projectnameSpace instance, providing integrity and confidentiality. 
Freshness leverages both local and remote \projectnameSpace, yielding all three guarantees. 

Figure~\ref{fig:combined_read} shows throughput, IOPS, and latency for these setups.
Starting with throughput, we observe how \projectnameSpace provides both integrity and freshness at a cost of just 1-2\%, reaching the size of network noise.
The lower initial throughput is caused by insufficient saturation of the protection pipeline.
As our latency increases for each request, we require higher throughput to hide it within a batch (e.g., encryption units on the BF3).
Throughput overheads are lowest as the proportion of time spent on checks in larger reads is smaller than in other cases.
IOPS for random and sequential patterns show a similar trend, where the protection pipeline requires 4-5 cores to match bare metal performance.
IOPS start with an overhead of 40-50\%, which decreases to 1-2\% for sequential reads and 5\% for random reads.
In both cases, freshness costs around 1\% performance overhead.
Single-core jobs averaged 6\% CPU usage with each additional core representing a 3 percent point increase, regardless of whether \projectnameSpace has been used.
These results are considerably better than existing methods, which, for sequential reads, can use up to 60\% of the CPU (Figure~\ref{fig:overheads}) while losing 12\% of performance and providing fewer guarantees than \projectname.
Similarly, for IOPS, at the same resource level, we achieve 5-6\% overheads instead of 30\%.
%

Latency overheads for integrity remain within 30\% and reach around 10\% at the higher percentiles.
The almost constant 9 µs overhead is mainly due to encryption.
For freshness, overheads grow earlier as additional checks are conducted on the data, such as comparing to the tree, and adding network freshness protections.
These can create contention, leading to a shift in the distribution.
Freshness behaves identically to integrity, except in the 50-80 percentile range, where the overheads increase by 32-56 µs. 

\vspace{-1em}
\subsection{Impact of metadata cache pollution (Q2)}

\begin{figure*}[b] 
    \centering
    \includegraphics[width=\textwidth]{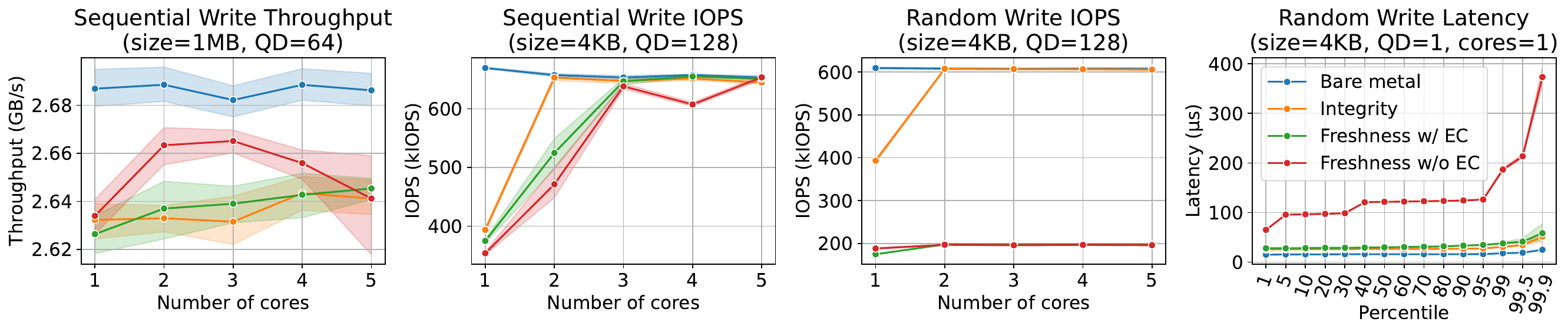} 
    \caption{Performance of \projectnameSpace across different write patterns.}
    \label{fig:combined_write}
\end{figure*}

In the previous experiment, we assumed an entirely fresh sector metadata cache for freshness.
However, this is not always the case.
Understanding how overhead scales with the degree of dirtiness of this metadata cache is critical.
Figure~\ref{fig:cache_pollution} shows how the performance of different patterns changes with the percentage of blocks that have a polluted cache with four cores.
Zero percent pollution implies a fresh cache, while 100\% pollution implies a system that gains no benefit from our metadata cache optimization.
We observe that throughput is not significantly affected by pollution and drops by only 1\%.
This behavior is expected, as requests typically require only one additional metadata sector read for many data reads.
However, both IOPS and mean latency are considerably affected.
For sequential reads, the main issue is IV cache contention, where multiple threads are serialized when accessing the corresponding cache entry.
For random reads, the issue is the metadata-to-data transfer ratio.
On average, each random read will require a write-back of the existing cache entry, a subsequent read of the correct IV sector to the cache, and then the actual read.
This also significantly increases latency when the cache is polluted.

\subsection{Writes and eventual consistency (Q1, Q3)}

Figure~\ref{fig:combined_write} shows write patterns.
In these, we also differentiate two freshness sets: one with (\verb|w/ EC|) and one without (\verb|w/o EC|) eventual consistency.
Throughput shows a similar pattern to reads, with integrity and freshness overheads reaching 2\% and 3\%, respectively.
Integrity IOPS reach top performance at just two cores and achieve average overheads of 1\% for both sequential and random writes.
Freshness IOPS reaches top performance with three cores for sequential writes, also achieving less than 1\% overhead.

Freshness IOPS struggle for random writes, where their overheads saturate at 66\%.
%
%
%
%
%
%
%
The overhead here is high for the same reason as dirty read IOPS: the metadata-to-data movement ratio.
Because for any tree update, we need all the neighbors' IVs, we need to, on average, write back the cache block, read the new cache block, and write the actual data.
For sequential patterns and large writes, this means we read the metadata IV sector into our in-memory cache once, then reuse it multiple times.
For a random pattern, there is little reuse.
As we conduct three writes/reads, this roughly cuts the bandwidth in three.
While we can hide additional latency in sequential writes, we cannot hide higher bandwidth usage in random writes.
For integrity, the latency overhead stays within 10-15 µs until the 99.5th percentile.

Writes with EC perform better than those without EC.
For IOPS, the performance difference reaches 8\% of bare metal, and the EC version is more stable.
For latency, it is hundreds of percent.
This is expected as placing the tree off the critical path hides these overheads.
The latency overhead of the EC version above bare metal starts at 13 µs and slowly climbs to 33 µs after the 50th percentile.
%


\subsection{Benchmarks and applications (Q4)}
\subsubsection{IO500}
\begin{figure*}[t] 
    \centering
    \includegraphics[width=\linewidth]{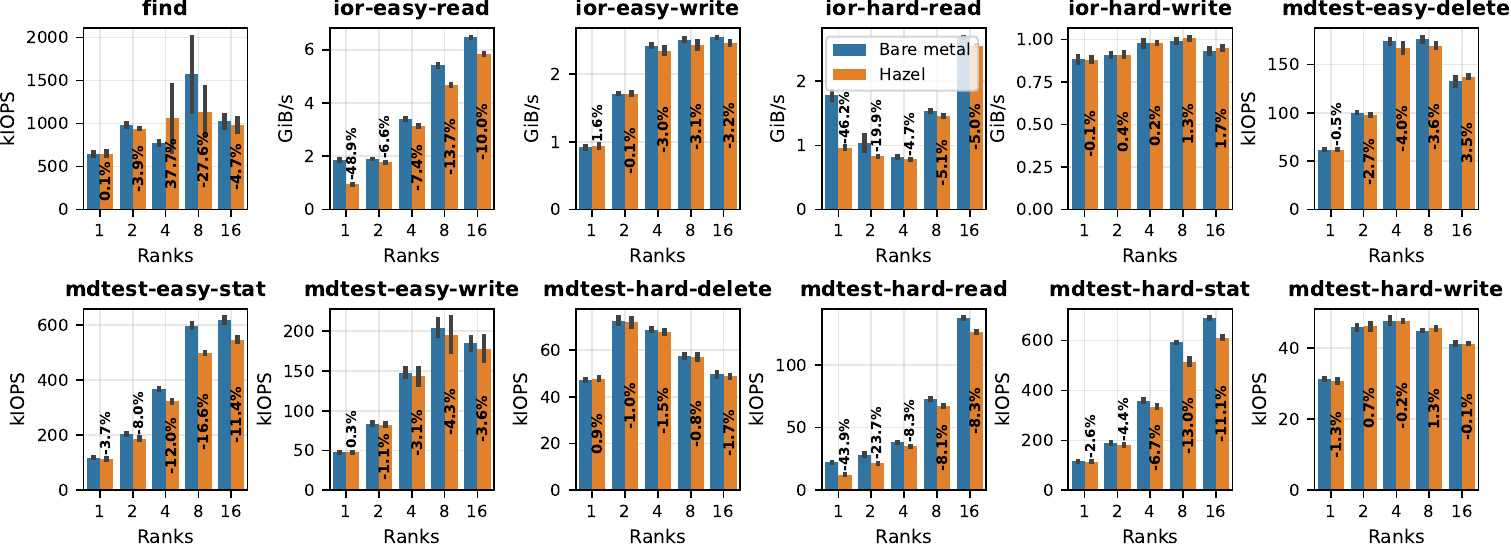} 
    \caption{Overheads for the IO500 benchmark suite where \projectname{} reaches 6.3\% overhead over all patterns and thread counts.}
    \label{fig:io500}
    \vspace{-0.5em}
\end{figure*}

\begin{figure*}[t] 
    \centering
    \vspace{-0.5em}
    \includegraphics[width=\linewidth]{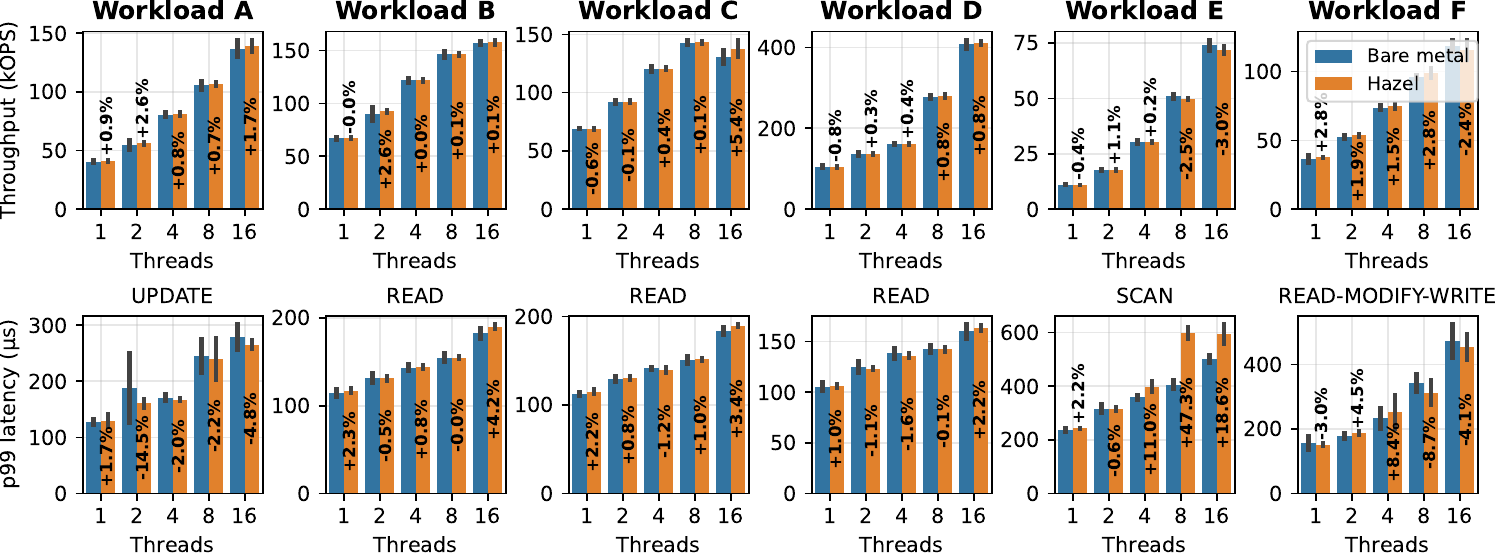} 
    \caption{Overheads for the YCSB benchmark deployed on RocksDB where \projectname{} reaches on average 2.2\% overhead for p99 latency and 0.6\% for throughput over all workloads and thread counts.}
    \label{fig:ycsb}
\end{figure*}

We deployed an \verb|ext4| file system on our setup and measured the performance of a well-known HPC benchmark: IO500~\cite{kunkel2016establishing}. 
IO500 evaluates a wide range of file system operations through microbenchmarks, stressing the underlying storage system.
To eliminate the effects of caching, we instructed IO500 to drop them during each experiment.

Figure~\ref{fig:io500} shows 12 different patterns belonging to the IO500 suite comparing bare metal and \projectname{} running with eventual consistency.
In most cases, the overhead of \projectname{} is below 10\%, averaging 6.3\%.
In some patterns, such as \verb|find|, the noise for 4-16 threads is large enough that \projectname{} remains within the performance of bare metal.
For the read-heavy cases such as \verb|ior-easy-read|, \verb|ior-hard-read|, and \verb|mdtest-hard-read|, we observe that at lower thread counts, the overheads can reach 40-50\%, then drop to around 10\%.
As with the synthetic results, for a low number of threads, the performance of small reads is bottlenecked by saturating the DPU cores, which cannot hide the latency overhead.
In this case, because the reads follow previous small writes, most in-sector metadata is dirty, which prevents the benefits of our fast-read path (in contrast to Section~\ref{sec:ml_training}).
The write and delete patterns show little overhead, around 1-2\%, with maximums reaching 4.3\%.
These results show that \projectname{} has low overheads across a wide range of workloads.

\subsubsection{YCSB}

To evaluate the performance of \projectname{} for a database system, we used the Yahoo Cloud Serving Benchmark (YCSB)~\cite{cooper2010benchmarking}.
We deployed YCSB with a key-value RocksDB~\cite{cao2020characterizing}.
This also serves as an example of a workload where the application's cache amortizes our overhead.
We configured RocksDB to have 10M records, each with a 1024B field, and ran YCSB over 5M operations.

Figure~\ref{fig:ycsb} shows the p99 and throughput performance across the six different workloads YCSB supports.
For the majority of workloads, \projectname{} achieves overheads within the noise, averaging a mere 0.6\% for throughput and 2.2\% for latency.
The only case where \projectname{} is considerably worse-performing than bare metal is for workload E, where the p99 latency increases significantly. 
This is predominantly because workload E comprises scans that randomly access the disk, which is where \projectname{} has large overheads due to low cache reuse.
With the addition of application caches, \projectname{}'s performance overheads can stay within system noise.

\subsubsection{Machine learning training}
\label{sec:ml_training}

Finally, we leveraged MLPerf's storage benchmark~\cite{balmauCharacterizingMachineLearning2022, MLPerfStorage} to evaluate the impact of \projectnameSpace on the performance of machine learning (ML) training pipelines, deployed in the past on disaggregated storage~\cite{zhuEfficientUserLevelStorage2019}
This is a key application for secure storage, given the confidential nature of datasets and the value of weights.
For our evaluation, we used ResNet50 and UNet3D, running over five epochs.
For the former, we simulated training using 16 NVIDIA H100 GPUs across 2557 data files, each 137 MB in size. 
For the latter, we simulated training on 1 NVIDIA H100 GPU using 3500 data files, each containing 140MB of samples.
%

\begin{figure}[] 
    \centering
    \includegraphics[width=\columnwidth]{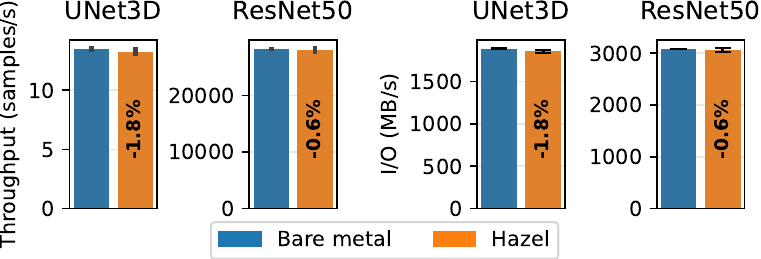} 
    \caption{Overheads for an ML training pipeline.}
    \label{fig:ml_overheads}
    \vspace{-1em}
\end{figure}

Figure~\ref{fig:ml_overheads} shows the performance of these models for the bare metal and freshness cases (with eventual consistency) across two metrics of interest: throughput of training samples, and the corresponding IOs.
The former represents the end-to-end practical impact, while the latter represents a storage impact.
The overhead of freshness remains at just 1-2\%.
As these workloads comprise predominantly large reads, they confirm the synthetic benchmarks in a practical deployment.
Note that such good performance is a direct result of our fast read path, as low write traffic over the existing data allows us to use fresh in-sector metadata.
As such, this workload is exactly where \projectname{} achieves best performance.

\section{Discussion}
Our evaluation leverages a modern, high-performance PCIe SSD, with a dedicated back-to-back link between the remote and local \projectname{} instances.
%
%
To contextualize our numbers in the scope of HPC systems, we compared them with some publicly available IO500 results~\cite{IO500SC25Production}.
Our testbed achieves \emph{per-node} performance similar to several reported systems (e.g., Sol, Alpha-Centauri, Lem). 
This is primarily because we did not experience network contention or use HDDs, SATA/SAS SSDs, or a parallel file system.
This allowed us to reach a representative end-to-end performance that these nodes provide.

We explore how \projectname{} scales as we add more high-performance disks.
Increasing their number does not modify the latency.
If concatenated, neither throughput nor IOPS increase.
Performance improves only when the disks are RAIDed, and increases linearly with RAID0 until the line rate is reached.
Local \projectname{} is bottlenecked by the BF3 GGAs, which are designed for line-rate.
Remote \projectname{} is bottlenecked by the HMT, which we have shown scales with sufficiently large writes.
However, as throughput and IOPS increase, the user will need more computation resources to drive the traffic.

%% file: 7_discussion.tex
\section{Related work}
\label{sec:related_work}

\hspace{1em}\emph{TDISP:} A key innovation enabling performance and threat model of \projectname{} is TEE Device Interface Security Protocol (TDISP)~\cite{tdisp2022}, based on a new PCIe extension: PCIe Integrity and Data Encryption (IDE).
TDISP enables attestation for PCIe devices, and IDE protects the transport to and from these devices.
These technologies enable us to extend trust to the local DPU in our work, which reduces resource usage by leveraging the DPU accelerators, and enables performance optimizations such as GPUDirect storage~\cite{GPUDirectStorageDirect2019}.
%
%



\paragraph{Secure cloud storage:} CSPs offer secure storage solutions in the form of full disk encryption~\cite{DataEncryptionOptions, DataEncryptionIntroduction, msmbaldwinOverviewManagedDisk}. 
These solutions are based on AES-XTS~\cite{dworkinRecommendationBlockCipher2010}, with the encryption tweak determined by the sector number.
Such implementation opens up chosen-ciphertext attacks~\cite{cipherleaks,cipherspaces}, as encrypting the same data at the same sector with the same tweak produces identical ciphertexts.
Furthermore, these solutions do not ensure integrity, freshness, or attestability.
%
%
%

\paragraph{CC storage:} Frameworks porting applications to TEEs typically provide storage solutions.
Gramine~\cite{tsaiGrapheneSGXPracticalLibrary2017}, Occlum~\cite{shenOcclumSecureEfficient2020}, and SCONE~\cite{arnautovSCONESecureLinux2016} protect by transparently encrypting and hashing files.
SecureFS~\cite{kumarSecureFSSecureFile2021}, DISKHIELD~\cite{ahnDISKSHIELDDataTamperResistant2020}, and Intel's IPFS~\cite{OverviewIntelProtected} create secure file systems specific to SGX. 
We considered approaching our design from the file system level, rather than the block device level.
However, we found that synchronizing protections at the disaggregated file-system level creates a larger scalability problem than using the block device layer.
%
%
Haven~\cite{baumannShieldingApplicationsUntrusted2015} provides protections on the block level in a similar fashion to \textit{dm-integrity}.
However, it does not handle freshness, is not TEE-agnostic, and is not fundamentally designed for disaggregated storage.

\paragraph{Integrity trees:} \textit{dm-x}~\cite{chakrabortiDmxProtectingVolumelevel2017}, Bonsai MTs~\cite{rogersUsingAddressIndependent2007}, and ProMT~\cite{alwadiProMTOptimizingIntegrity2021a} propose different MTs to protect local storage. 
Integrity trees have also been used for securing non-volatile memory~\cite{freijPersistLevelParallelism2020} and volatile memory in TEEs such as SGX~\cite{gueron:a_memory_encryption_engine_2016}.
VAULT~\cite{taassoriVAULTReducingPaging2018a} introduces a new type of tree organization that is more compact and minimizes depth.
Counters can be used to achieve a similar performance improvement~\cite{saileshwarMorphableCountersEnabling2018}.
EnclaveDB~\cite{priebeEnclaveDBSecureDatabase2018} uses an MT for database log freshness. 
These works do not address \projectname's challenges with CC-disaggregated, large-scale, long-term storage.

\paragraph{Distributed storage:} Works such as LustreFS~\cite{Lustre}, GPFS~\cite{schmuck2002gpfs}, HDFS~\cite{borthakur2008hdfs}, Ceph~\cite{weil2006ceph}, and GFS~\cite{mckusick2009gfs} provide distributed storage scalable to thousands of nodes.
Others like Octopus~\cite{luOctopusRDMAenabledDistributed2017}, Clover~\cite{tsaiDisaggregatingPersistentMemory2020}, Assise~\cite{andersonAssisePerformanceAvailability2020}, and Orion~\cite{yangOrionDistributedFile2019} optimize the performance.
Some works also focus on storage sNIC offloading. 
LineFS~\cite{kimLineFSEfficientSmartNIC2021a} introduces distributed file system offloading with sNIC pipelining.
NVIDIA SNAP~\cite{SNAP} virtualizes NVMes through hardware acceleration.
Gimbal~\cite{minGimbalEnablingMultitenant2021} optimizes the fairness of disaggregated storage.
None of these works focuses on secure storage with freshness.

\section{Conclusions}
We presented \projectnameSpace, a first-of-a-kind secure disaggregated storage solution.
We introduce a control path that disaggregates freshness checks and scales to PB-scale disks, and modify the data path to introduce novel concepts, such as a metadata cache and eventual consistency.
These enable \projectnameSpace to achieve overheads as little as 2\% on example applications while providing confidentiality, integrity, and freshness.
Combined with our portable SPDK implementation and BF3 support, we believe \projectnameSpace is a fundamental step for the future of scalable CC in supercomputers and data center clusters.